\documentclass{article}

\usepackage{arxiv2}

\usepackage[utf8]{inputenc} 
\usepackage[T1]{fontenc}    
\usepackage{hyperref}       
\usepackage{url}            
\usepackage{booktabs}       
\usepackage{amsfonts}       
\usepackage{nicefrac}       
\usepackage{microtype}      
\usepackage{lipsum}
\usepackage{graphicx}

\usepackage{amsmath}

\graphicspath{ {./images/} }

\usepackage{pagenote}
\usepackage{enumerate}

\usepackage{url}

\title{The sensitivity of the Deffuant model to measurement error}

\author{
Dino Carpentras\textsuperscript{1,2}
  \And
Michael Quayle\textsuperscript{1,3}
  \And
  \\
\textsuperscript{1}Social Dynamics Lab, Department of Psychology, \\ Centre for Social Issues Research, University of Limerick, Ireland \\
\textsuperscript{2}MACSI (Mathematics Applications Consortium for Science and Industry), \\ Department of Mathematics and Statistics,  University of Limerick, Ireland \\
\textsuperscript{3}Department of Psychology, School of Applied Human Sciences,\\ University of KwaZulu‐Natal, Scottsville, South Africa
}

\usepackage{natbib}
	\setcitestyle{authoryear,round,aysep={}}
	
\begin{document}
\maketitle
\begin{abstract}
Opinion dynamics models have an enormous potential for studying current phenomena such as vaccine hesitancy. Unfortunately, to date, most of the models have little to no empirical validation.
One major problem in testing these models against real-world data relates to the difficulties in measuring opinions in ways that map directly to representations in models. Indeed, this kind of measurement is complex in nature and presents more types of measurement error than just classical random noise. Thus, it is crucial to know how these different error types may affect the model’s predictions. In this work, we analyze this relationship in the Deffuant model.
Starting from the psychometrics literature, we first discuss how opinion measurements are affected by three types of errors: random noise, binning, and distortions (i.e. uneven intervals between scale points). While the first two are known to most of the scientific community, the third one is mostly unknown outside psychometrics. Because of that, we highlight the nature and peculiarities of each of these measurement errors.
By simulating these types of error, we show that the Deffuant model is robust to binning but not to noise and distortions. Indeed, if a scale has 4 or more points (like most self-report scales), binning has almost no effect on the final predictions. However, prediction error increases almost linearly with random noise, up to a maximum error of 40 \%. After reaching this value, increasing the amount of noise does not worsen the prediction. Distortions are most problematic, reaching a maximum prediction error of 80 \%.
Up to now most of the research focused on the properties of the models without analyzing the types of data they may be used with. Here we show that when studying a model, we should also analyze its robustness to these types of measurement error.

\end{abstract}








\section{Introduction}

Opinion dynamics is a sub-field of agent-based modelling focused on people’s opinions \citep{Flache, Castellano_2009}. Models from this field, are good candidates for better understanding pressing social problems, such as vaccine hesitancy. However, current opinion dynamics models are rarely used with real world data \citep{Flache, Duggins, Jia, Innes}. This means that, currently, it is not possible to tell if the prediction of a model, such as the Deffuant one, will be accurate or not.

Indeed, since 2010 different articles and reviews exhorted researchers to validate their model against real-world data \citep{Flache, Castellano_2009, Dong, Valori}. Probably because of that, recently, some projects aimed at achieving this goal received substantial funding from different agencies \citep{ToRealSim, Dynamod}. Thus, we expect that in the following years more and more effort will be dedicated in validating opinion dynamics models.

One of the biggest issues in bridging this gap between models and data, lays in the fact that measuring opinions is a complex process. Indeed, physical measurements rely on well-defined units of measurements, while nothing similar exists for measuring opinions \citep{Tversky1, Stevens_1946}. Notice that this does not mean that it is not possible to make such a measurement. Indeed, the field of psychometrics has developed theory, methodology and standards just for this type of task \citep{Stevens51, DeVellis, IRT}. However, the types of measurement error that affect opinions are not the same that affect physical measurements.

In this work, we want to start bridging the gap between opinion dynamics models and real-world data by analyzing how the Deffuant model is sensitive to different types of measurement error. We choose this model both for its impact on the field and for its simplicity \citep{Deffuant2000, Flache}. However, our results show that this type of analysis should be carried out before seeding any agent-based models with real psychometric data.

In the next section we discuss how psychometrics theory introduces three types of measurement errors. In the method section, we discuss how we simulated each one of them and estimated their effect. In the analysis section we show, via simulations, how each type of measurement error impacts the Deffuant models’ predictions. Finally, in the discussion, we clarify the implications of these results and how researcher may avoid, or at least, minimize measurement-related prediction error.

\section{Psychometric measurements}

\subsection{A review of measurement errors}

Before testing the Deffuant model against measurement errors, it is important to clarify the effects that influence a measurement. A common approach in the so-called “hard sciences” is to suppose the existence of a true value $v$ to be measured \citep{Taylor, Halliday}. This value is usually considered to belong to the real numbers. As the existence of $v$ is a controversial assumption in psychology, in the next sub-section we will better discuss this hypothesis and how the following analysis can be adapted to a case in which no real value exists. However, for simplicity, we start by assuming the existence of such a value. Furthermore, we start by analyzing each type of noise independently.

When we perform a measurement $m(v)$ affected by random noise we obtain as result:

\begin{equation}
\label{e1}
m(v) = v + r
\end{equation}

Where $r$ is a randomly distributed variable representing the noise. It is also commonly supposed that $r$ has a normal (also known as “Gaussian”) distribution of mean zero and a standard deviation $\sigma$ \citep{Taylor}. Thus $\sigma$ can be used as a measurement of the noise’s intensity.

When the measurement instrument has a relatively small number of levels it is common to include a quantization (or discretization) error \citep{Sayood}. For example, an instrument measuring voltage may be able to display either 5 or 6 V, but not the values in between. Similarly, a 10-point self-report measure political opinion would be able to distinguish between a person with opinion 1 to a person with opinion 2, but not between people within this range. Meaning that a person with opinion 1.6 will probably receive a score of 2, making this value indistinguishable from the score of the second participant. In general, we will write this type of measurements as:

\begin{equation}
\label{e2}
m(v) = bin(v)
\end{equation}

Where $bin()$ is a rounding-like function, such as round. For example, we can suppose that $v$ is a number between 0 and 1 and that we are measuring it on an 11-point scale from 0 to 10. Then, the bin function will take the form of $bin(v)= round(10*v)$. 

Up to now, we were able to draw some parallel with measurements in physics. However, there is another fundamental effect which is unique to measurement of variables such as opinions. Indeed, in physics, measurements are performed in specific units (e.g. meters or feet for measuring distance). Unfortunately, in psychometrics, there is no things such as a unit of opinion. This does not mean that we will not be able to measure opinions. Indeed, this is almost a standard procedure in the literature. However, our measurement will not be linear with respect to the true value \citep{Tversky1, Tversky2, Tversky3} even if we suppose that such a true value exists. Thus, we can represent the measurement process as:

\begin{equation}
\label{e3}
m(v) = h(v)
\end{equation}

Where $h()$ is a monotonic function, which, in general will be non-linear. Where a function f is considered to be linear if for every scalar $\alpha$ we have $f(\alpha v)= \alpha f(v)$.

This non-linear kind of measurements are often referred as “ordinal measurement” \citep{Tversky1} as they preserve order. Indeed, if there is an order in the true values, such as $v_1<v_2$ then the same order is preserved in the measurement: $h(v_1 )<h(v_2 )$. 

However, notice that this kind of measurement does not preserve distance and proportions. For example, if $v_2=2 v_1$ then in standard measurement we would expect also $m(v_2 )=2 m(v_1)$. However, since $h$ is a non-linear function, this relationship would not be true.

Because of this feature, this kind of scales have been often graphically represented as a ruler with uneven spacing between the ticks. With such a ruler, we could say that an object of length 6 is longer than an object of length 3, but not that it is actually twice as long.

Many people here may be confused from the fact that $2*3=6$, thus, they think that one object should be twice as long as the other. However, to reason correctly, we should differentiate between the object itself (or true value) and its measurement in this distorted scale. Indeed, in this case we can say that one measurement is twice the other, even if this relationship does not exist between the objects. In formula: $m(v_2 )=2 m(v_1)$ but $v_2 \neq 2 v_1$.

Another small difference between measuring opinions and measurement in physics is that psychometric scales are usually based on several items (i.e. questions). This means that a single measurement $m(v)$ is usually obtained by combining responses to several items from the same person. However, for simplicity, here we will just focus on the scale measurement errors, without analyzing how such a scale has been obtained.

While the first two types of errors are well known even in classical physics, the problem of distortions become particularly relevant with psychometric measurements. Also because of this, there are many studies showing how these distortions may affect standard procedures in psychological measurement, such as in t-test and ANOVA procedures \citep{Lanz, Glass, Feir, Baggaley}. However, to our knowledge, almost nothing is known on how they may impact agent-based modelling.

\subsection{On the existence of the true value}

In our argument so far we have assumed the existence of a true value, similarly to classical physics. This assumption has already been discarded in quantum physics, where due to Heisenberg uncertainty principle it is not possible to identify a true value for measurements such as position or speed \citep{Heisenberg, Landau}.

Something similar happens also in psychometrics where we cannot measure $v$ but only $h(v)$ \citep{Tversky1}. Furthermore, $h$ is an unknown function. This means that we cannot measure it and, even if we try to use another measurement, we will only end up with another unknown transformation. For example, two measurements of the same value $v$ on two different scales will result in  $h_1 (v)$ and $h_2 (v)$. Thus, even by adding more measurements, the system will remain underdetermined, and we would not be able to find the true value of $v$ or, equivalently, the shape of $h$.

This has brought many to the conclusion that a true value may not even exist as a numerically quantifiable internal construct \citep{Howarth}. Others, instead, suppose its existence and that it should be theoretically possible to obtain it though procedures based on a theory called additive conjoint theory \citep{Tversky1, Luce64}. In both cases, researchers agree that is practically impossible to extract a true value from most of the available data \citep{Heene, Borsboom}.

However, the absence of a true value does not mean that we cannot use the measurement model provided in our previous equations. Indeed, while we cannot compare scales to a true value, we can still compare scales to an arbitrary reference scale without any additional claims about its properties. 

For example, for measuring political opinion we could use the “12 Item Social and Economic Conservatism Scale” (SECS) \citep{Everett}. Alternatively, we can choose any other scale from the literature or even create our own scale, as long as it is measuring the right construct (in this case “political opinion”). This is similar to the situation in physics, where we can choose either meters or feet, or any other unit as long as it is a good measurement of distance.

Thus, in the previous equations we can just replace the true value $v$ with $v_r$, the value on a reference scale. Notice also that while the previous assumption was extremely demanding, the new assumption is always met in practice. Indeed, initially we assumed the existence of a true value, while now we just assume that we can generate a scale. This process is routinely done in psychology.

Notice also that this choice of the reference scale is useful also for identifying $h$. Indeed, while we cannot measure the relationship between a scale and the true value (as it will require to know the true value) we can always compare two scales. 

For example, in figure 1 we see the comparison of two scales of trust in science. Each one has been obtained averaging three questions on trust in science from the Wellcome Global Monitor \citep{WGM}. This kind of flexible item selection is common in social sciences.  Since the original survey contained only five questions on this topic the two scales have one of the three items in common. 

From the plot, we can see that people that scored 6 on scale 1 also scored an average of 6 on scale 2. However, people who scored 10 on scale 1 scored on average 7 on scale 2.  Thus, this plot practically shows us that we can choose a reference scale (scale 1) and compare it to another one (scale 2). Furthermore, it also graphically shows us the three errors we just discussed. Indeed, the error bar shows the effect of random noise, while the fact that points do not follow a linear behavior (orange dashed line) shows us the effect of distortions. Finally, the fact that we have only 10 points (from a score of 3 to 12) shows us the fact that both scales have only a few levels (i.e. binning) instead of being a continuous variable.

\begin{figure}[!t]
\centering
\includegraphics[width=0.5\textwidth]{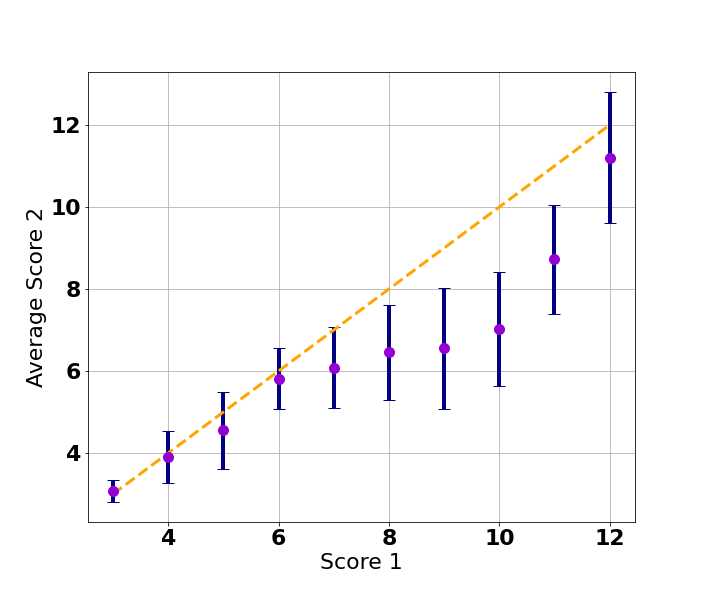}
\caption{Average score on scale 2 versus the score on scale 1. The dashed line shows the ideal linear behavior. The size of the error bar is due to random noise and the non-linearity shows us the effect of distortions.}
\label{X1}
\end{figure} 

\section{Method}
\subsection{The Deffuant model and parameters}

As mentioned before, the Deffuant model \citep{Deffuant2000} is a simple model which in general leads to a simple dynamic behavior. Every agent’s opinion is represented as a real number between 0 and 1. In this way, it is possible to calculate the distance between their opinion. If their distance is higher than the parameter $\varepsilon$ they will not be able to interact. Otherwise, they will move towards the average opinion. The amount of movement is proportional to the parameter $\mu$ and it follows the formula:

\begin{equation}
\label{e4}
o_i(t + 1) = (\mu - 1)  o_i(t) + \mu  o_j (t)
\end{equation}

Where $o_i (t)$ is the opinion of agent $i$ at time $t$.

For the dynamic evolution of the system, couples of agents are randomly chosen at every round. Here we use $T$ to denote the maximum number of rounds. 

Several works focused on the convergence state of the Deffuant model. However, as with many opinion dynamics models, the Deffuant model converges to an unrealistic distribution. Indeed, when the model has converged, all the people share the same $N$ opinions, where $N$ is roughly $1/\varepsilon$. 

This means that the Deffuant model may be useful not for predicting the final state of an opinion system (which in the real world is usually not static) but for making more short-time predictions. This is similar to what happens in meteorology, where models make bigger and bigger prediction errors as they try to predict further in the future. Thus they are mostly used for short-term predictions and they are constantly fed with the latest data \citep{Lean}.

Because of this, we did not focus on the predictions of the converged model, but on the predictions at shorter time intervals. Furthermore, as the convergence does not only depend on T but also on the number of agents, we usually represent time using $T_r=\frac{T}{N_a}$ , where $N_a$ is the number of agents.

Besides $\varepsilon$, $\mu$ and $T_r$ there is another “parameter” which is often not discussed in the literature: the initial distribution. Indeed, a common assumption is to start simulations using a uniform distribution. However, this is not a good representation of real-world opinion distributions, which can have more complex shapes and properties that can affect the model dynamics \citep{Valori}. Thus, since we are interested in the effect of real-word distributions, we simulated distributions generated by using a third order interpolation of 4 randomly generated numbers. Two examples of these distributions are shown in figure 1.

Despite not being included in the original article from Deffuant and Weisbuch, another part of the system which is often studied is the network topology \citep{Fennell}. However, since our analysis already depends on several parameters, we choose to not introduce an extra one. Thus, all the dynamic behavior we will discuss in the following will be on a fully-connected network (i.e. everyone can interact with everyone else, subject to the bounded confidence interval $\varepsilon$).

\begin{figure}[!t]
\centering
\includegraphics[width=0.8\textwidth]{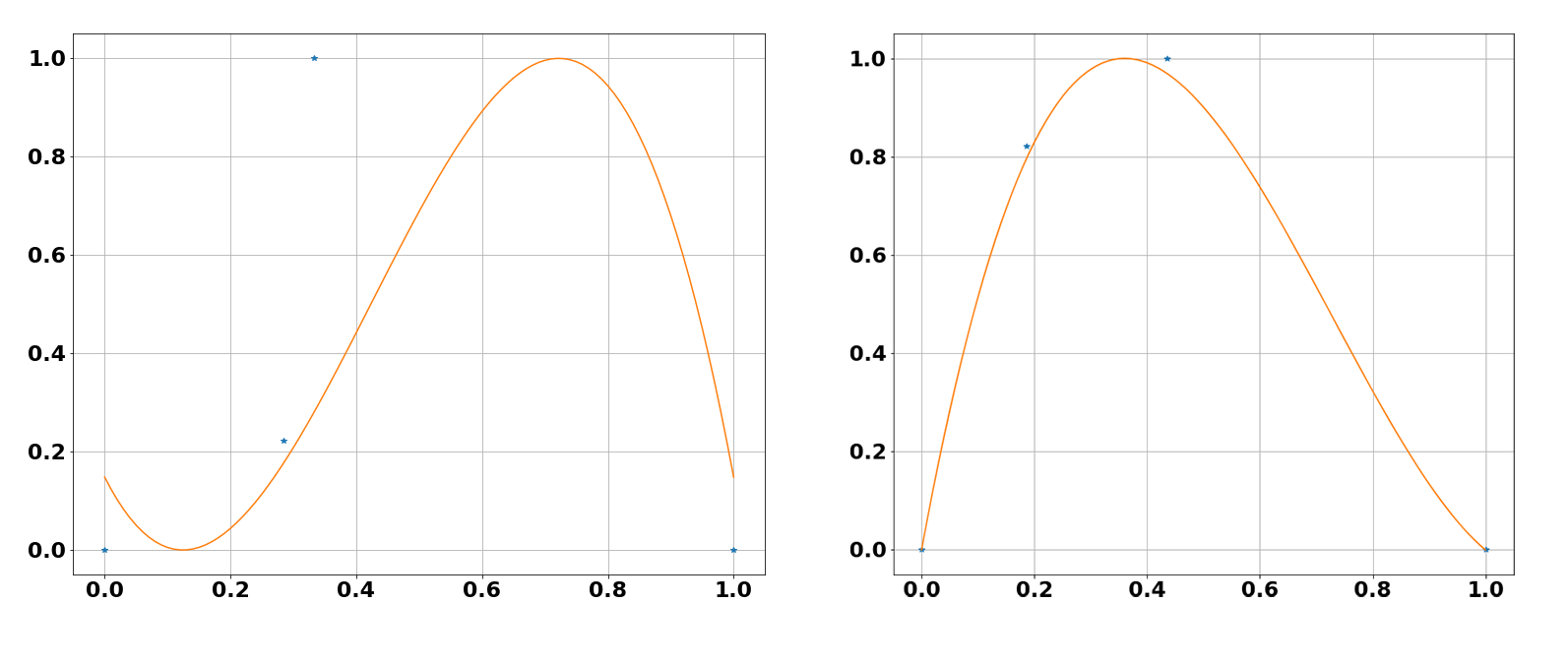}
\caption{Two examples of distributions used to seed the model.}
\label{X2}
\end{figure} 

\subsection{Management of the parameter space}

Despite the Deffuant model being one of the simplest models in the opinion dynamics literature, we can see that our analysis has to deal with several parameters. Because of this, a full exploration of the parameters space is not feasible. A solution used in some works consists in fixing $T$, $N_a$, $\varepsilon$ and $\mu$ to some specific values for the entire analysis \citep{Fennell}. Another common approach in these cases is the so called OFAT method (one factor at the time) \citep{Broeke}. This method consists in fixing all the parameter to a set of chosen values and varying only one of them at the time. Once the exploration of that parameter has been completed, it is fixed to its original value and a new parameter is explored.

Since we want our results to be as general as possible, we used a mixed method. Indeed, as we will discuss in the next section, the prediction error appears to be independent on $T_r$ and $N_a$. Because of that, the analysis will be carried out with $N_a=200$ and $T_r=1$. Since the other parameters have an effect on the prediction error, we decided to randomize them for each run. Specifically, we randomly choose $\varepsilon$ and $\mu$ between 0 and 0.5 as these are the standard extremes used for analyzing the Deffuant model. Instead, the distribution curve is initialized each time in the way discussed in the previous section. Furthermore, we also set all the error sources to zero, besides the one under analysis. 

For example, if we are studying the effect of binning, we will set the scale to a specific value (e.g. a 10-point scale) and set the noise and distortions to 0. Then, we will randomly choose $\varepsilon$, $\mu$ and the distribution curve and run the simulation. Since there are several random parameters in the simulation, we will repeat this process many times, by selecting again a 10-point scale, but new random values of $\varepsilon$, $\mu$ and the distribution curve.

\subsection{Measurement of error and distortions}

We calculated the prediction error by using Theil’s inequality coefficient \citep{Leuthold}. This value has been employed to measure how similar two predictions are. Indeed, for two predictions X and Y it can be calculated as:

\begin{equation}
\label{e5}
E(X,Y) =  \frac{\sum_i \sqrt{x_i^2 - y_i^2} }{\sum_i (\sqrt{x_i^2} + \sqrt{y_i^2}) } 
\end{equation}

Where $x_i$ is the opinion of person $i$ in prediction $X$. Notice that this value is always between 0 and 1. The value 0 represents the case of perfect prediction (i.e. $x_i=y_i$) while 1 represents the case of maximum error. This is achieved if either $X$ or $Y$ is constantly zero.

In our analysis $X$ is the prediction we obtained from the simulation without measurement errors, while $Y$ is the prediction obtained by including the measurement error under analysis. This means that to generate $X$ and $Y$ we run the Deffuant model using the same parameters (initial distribution, $T$, $N_a$, $\varepsilon$ and $\mu$). The only difference is that to produce $Y$ we also added a measurement error before running the model. Thus $E(X,Y)$ represents the prediction error we would observe comparing two scales: one with measurement error and another without.

As reference scale we chose a 101-point one as precision of self-report measures rarely exceeds this. Indeed, while any scale could be used as a reference, for this study we selected one that would be similar to a maximally precise psychometric measurement under ideal conditions and to then simulate what happens as precision is decreased.

Regarding the simulations, we chose not to limit the precision of the model dynamics. Meaning that only the initial values of seeded data would be represented on the scale levels, but not the intermediate dynamics or final level. To provide an example, we can consider having an opinion scale with only two points. Thus, the model would be seeded with all agents either having an initial opinion of 0 or 1. However, during the simulation, they will still follow the Deffuant dynamics without being limited to these two values (e.g. they could move to 0.6 even if this is not a point on the initial scale). These final values will be then used to calculate Theil’s coefficient.

As we just discussed, the amount of binning is quantified by the number of points on the measurement scale. The amount of random noise, instead, is quantified by $\sigma$, the standard deviation of the normally distributed noise variable. Finally, distortions are quantified by the deviation from the linear behavior:

\begin{equation}
\label{e6}
D(X,h) =  \frac{\sum_i |x_i - h(x_i)| }{\sum_i x_i } 
\end{equation}

Where $x_i$ is the set of values on the reference scale $h(x_i)$ are the equivalent values on the measurement scale. Similarly to the Theil’s coefficient, $D$ ranges between 0 and 1. Specifically, 0 would be the case of perfect linearity, while 1 would be the case of maximum non-linearity.

\section{Analysis}

\subsection{Intrinsic stochastic error}

Before studying how different measurement error affect the prediction error, we should first notice that the Deffuant model is affected by random variations. Even if their effect tends to disappear when the model converges, they may still have an effect for predictions made before that point.

An extremely simple example would be to make predictions using $T=1$. In the first run of the model, we could have that person 1 and person 10 are randomly chosen for the interaction. However, if we repeat the same configuration for a second run, we could choose a different couple. This means that predictions would be different even without introducing any measurement error. We refer to this effect as the “intrinsic stochastic error.”

We measured the intrinsic stochastic error by randomizing $N_a$ between 100 and 1000, $T_r$ between 0.1 and 5, $\varepsilon$ and $\mu$ between 0 and 0.5. We observed that the average error to be 0.054 with a standard deviation of 0.038.

As we can see from figure 2, the intrinsic stochastic error is not dependent on $T_r$ and $N_a$. This is the reason why for further analysis these two parameters will be fixed to $T_r=1$ and $N_a=200$. We choose these two values as they are small enough for efficient computation.

Regarding $\varepsilon$, we observe that it influences the prediction error only for $\varepsilon < 0.1$, while above this value the error keeps oscillating around its mean value. Instead, regarding $\mu$ the error keeps increasing as this value increases. It is important to notice that $\mu$ has a direct influence on how much each agent will move during the simulation. Thus, it makes sense that the higher this value the bigger the error will be in the final prediction. Similarly, $\varepsilon$ has an indirect effect on the agents movement. Indeed, for very small $\varepsilon$ agents will be able to interact only with their direct neighbors. 

\begin{figure}[!t]
\centering
\includegraphics[width=0.8\textwidth]{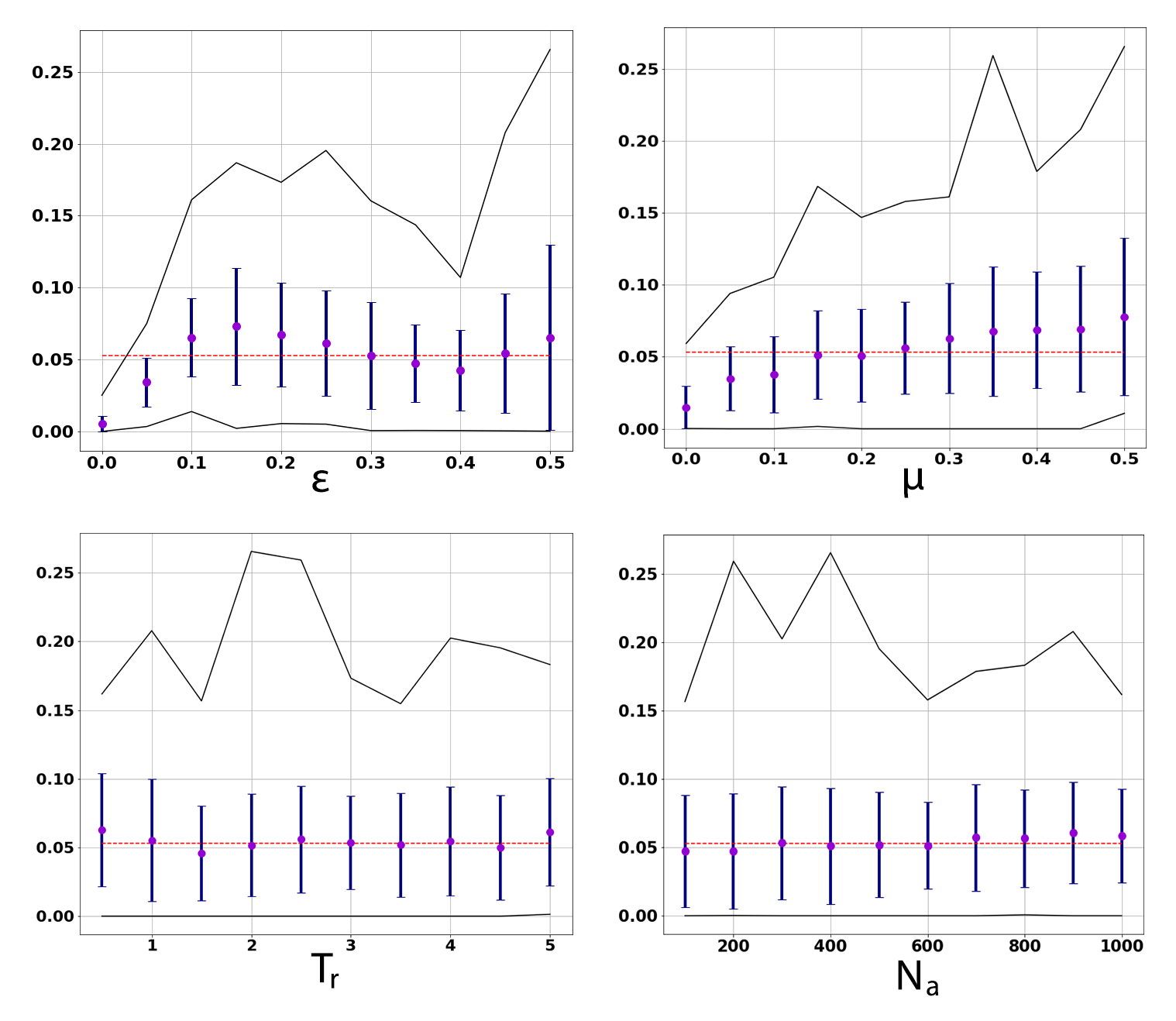}
\caption{Prediction error depending on different parameters. Black lines represents the maximum and minimum errors. Red dashed line represents the intrinsic stochastic error. The purple dot represents the average error value, while the error bar represents the standard deviation of the error.}
\label{X3}
\end{figure} 

\subsection{The effect of random noise}

The first measurement error that we analyze is the random noise. As mentioned before, this is usually modeled as normally distributed with mean of 0 and a standard deviation of $\sigma$. As measurement errors affect only the initial data, it is interesting to see what its effect on the initial distribution is.

To better understand how this noise works, we can think of an example in which every agent on the reference scale has an opinion of 0.5. If now we introduce the randomly distributed noise, we will have that the mean opinion on the new scale will still be 0.5, however agents’ opinions are spread around this value following a normal distribution with a standard deviation of $\sigma$. 

If the initial distribution is not just a single peak, but a more complex shape, we can still use the previous procedure to see where agents will be. Indeed, also in this case, all agents which initially were in 0.5 will be randomly distributed around this value. All the agents which were in 0.6 will be randomly distributed around 0.6, and so on for each position.

Mathematically, this operation is called convolution \citep{Smith} and it is represented as:

\begin{equation}
\label{e7}
f_1(x) = f_0(x) * n(x)
\end{equation}

Where $f_0$ is the initial distribution, $f_1$ the final one, $n$ is the noise function and $*$ is the symbol of convolution.

Connecting this type of noise to convolution has the advantage that this mathematical operation has been intensively studied and its effect on other functions is well known. Indeed, convolving a function with a normally distributed one, has the effect of smoothening the first function. For example, in optics, a blurred image is modeled as the convolution of the high-quality image with a normally distributed function \citep{Soskind}.

In figure 4 we can see the effect of the random noise on the prediction error. As expected for 0 noise, the prediction error is still close to the intrinsic stochastic error, which is the case of no measurement error. As the amount of noise increases, the prediction error keeps linearly increasing up to a value of roughly 40 \% for $\sigma = 0.5$. After this value of $\sigma$, the prediction error does not seem to increase anymore.

We can then approximate the noise with a piecewise linear function as:

\begin{equation}
    \begin{cases}
      0.05 + 0.70\sigma \qquad if \; \sigma \leq 0.5\\
      0.4 \qquad \qquad \qquad \, if \; \sigma > 0.5
    \end{cases}
\end{equation}

Notice that while we analyzed the effects of noise up to $\sigma = 1$, usually in empirical data the amount of noise will be much smaller. For example, we can use the data used for producing figure 1 (i.e. from the Wellcome Global Monitor) to calculate that between the two scales there is a measurement noise of roughly 10 \%. By substituting this value into the previous formula, we have that the average prediction error would be 8.5 \%, which is a relatively small increase to the intrinsic stochastic error.

\begin{figure}[!t]
\centering
\includegraphics[width=0.5\textwidth]{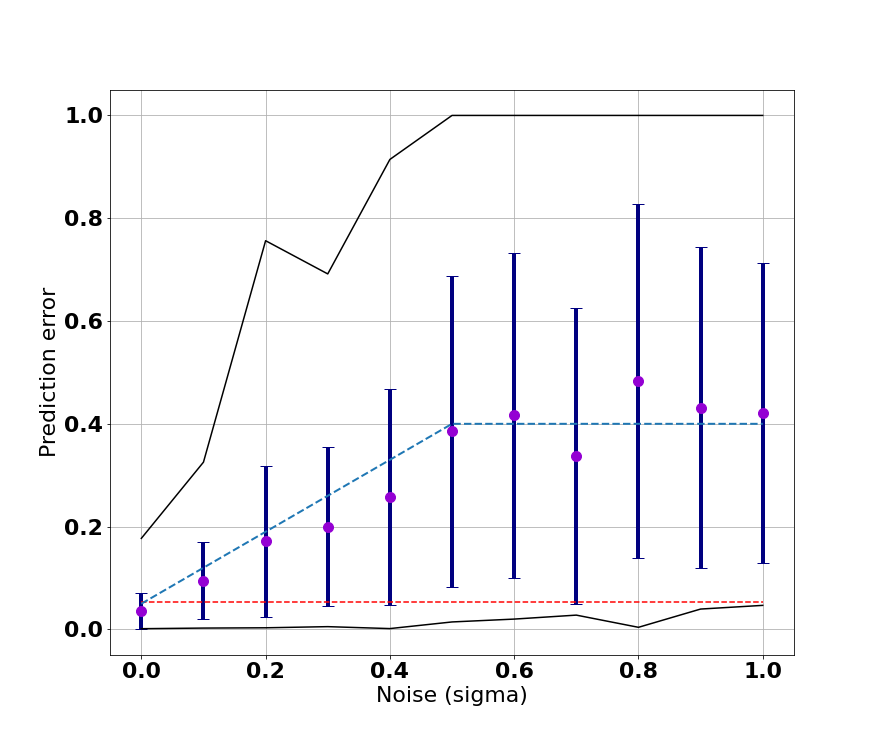}
\caption{Prediction error versus the amount of measurement noise measured in terms of $\sigma$. Black lines represent the maximum and minimum errors. Red dashed line represents the intrinsic stochastic error. The purple dot represents the average error value, the error bar represents the standard deviation of the error. The blue dashed line is the piecewise linear approximation.}
\label{X4}
\end{figure} 

\subsection{The effect of binning}

The second effect we study is due to binning. This is a fundamental effect as measurements in psychometrics usually have a few levels, as opposed to measurements in physics which may easily have thousands or even millions of levels. Indeed, in psychometrics it is common to measure someone’s opinion on a topic by asking one or a few questions. Each question is also usually limited to a few response options, typically ranging from 2 to 10. The answers from single questions are then combined to produce an overall score with increased number of levels \citep{Nunnally}. As previously mentioned, the reference scale in our simulations has 101 levels, so it is interesting to analyze how prediction error changes as we decrease the number of levels on the new scale.

As we can see in figure 5, the error is almost constant for scales with 4 or more levels. Then the error has a very rapid increase below 4. The worst situation is, of course, the case of only 2 levels, which produces an average prediction error of 34\%. 

This result is extremely positive for people intending to use the Deffuant model with real-world data. Indeed, most measurement items provide more than 3 levels, thus providing the minimum prediction error. 

Notice also that in this specific case we do not provide a piecewise linear approximation as the prediction error is non-constant only in two points. Thus, it makes more sense to directly report their value, instead of providing an approximation. Thus, the prediction error for 2 and 3 levels is respectively 34\% and 12\%.

This analysis shows also some kind of data-incompatibility between the Deffuant model and some other models. Indeed, literature in agent-based modelling is often interested in the difference between models \citep{Coates}. One sub-class of models in opinion dynamics is based on binary states, such as in the voter model \citep{Holley, Vazquez}. This means that this class of models is designed for dealing with opinions measured on 2-point scales. As showed, when the Deffuant model is used with this type of data it is affected by extremely high prediction uncertainty. Meaning that data which would be perfect for one  model, would produce terrible results in another.

\begin{figure}[!t]
\centering
\includegraphics[width=0.5\textwidth]{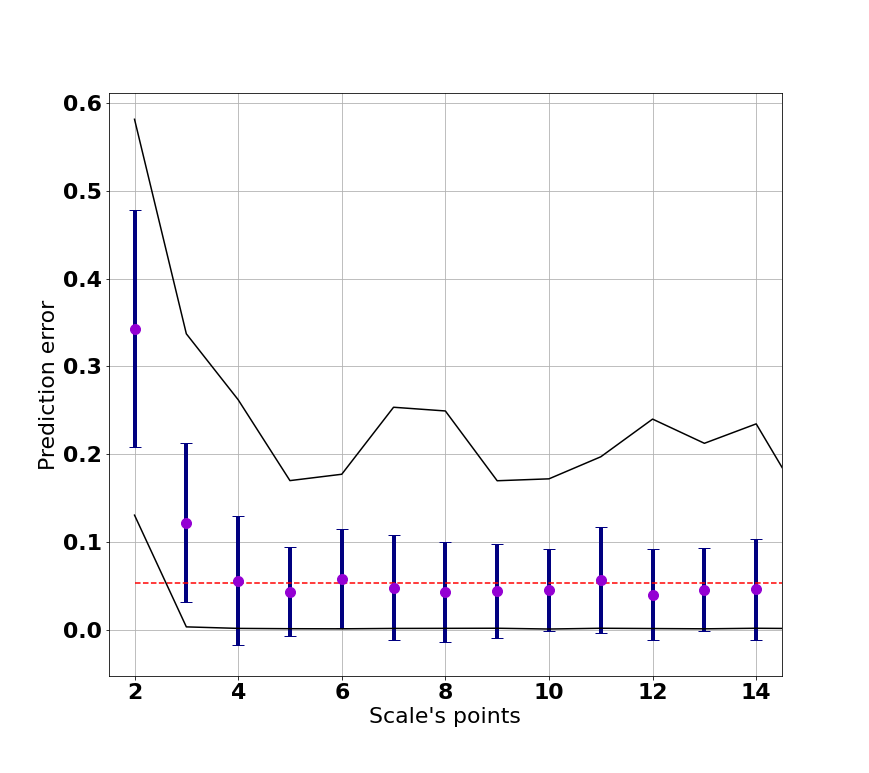}
\caption{Prediction error versus the number of points in the scale. Black lines represent the maximum and minimum errors. Red dashed line represents the intrinsic stochastic error. The purple dot represents the average error value, the error bar represents the standard deviation of the error. }
\label{X5}
\end{figure}

\subsection{The effect of distortions}

The final analysis regards the effect of distortions. As mentioned in the introduction, these non-linear transformations are unavoidable features of psychometric measures such as Likert-type scales which are the standard for measuring opinions \citep{Nunnally}. From figure 6 we can see that distortions up to 0.1 produce almost no increase in the prediction error. However, after this value, the prediction error starts rising almost linearly with the distortion value. This effect is so strong to reach an average prediction error higher than 80 \%. Following this behavior, we approximated these data with the following piecewise linear function:

\begin{equation}
    \begin{cases}
      0.05 \qquad \qquad \quad  \; if \; D \leq 0.1\\
      0.8 D - 0.03  \qquad  if \; D > 0.1
    \end{cases}
\end{equation}

To have a better idea of how strong this effect may be in real-world data, we can use again the data form the Wellcome Global Monitor. Here we find a distortion value of 0.26, which leads to an average prediction error of 18 \%. This is more than 3 times bigger than the intrinsic stochastic error.

To better understand this result, let us suppose two people decides independently to use the data from the Wellcome Global Monitor with the Deffuant model to predict trust in science. If one person uses scale 1 and the other uses scale 2, then their predictions will differ by an average of 18 \%.

This represents an important factor for using real-world data with the Deffuant model. Indeed, while the noise and the number of levels have a quite limited effect on the prediction error, distortions may easily produce much higher prediction errors.

\begin{figure}[!t]
\centering
\includegraphics[width=0.5\textwidth]{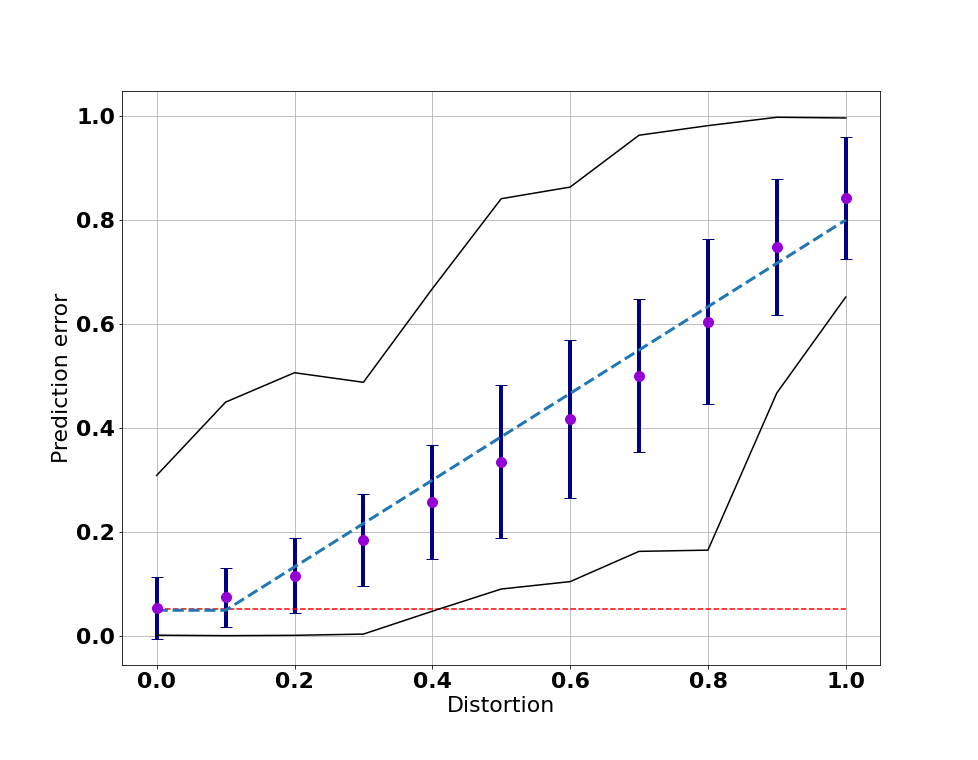}
\caption{Prediction error versus the distortions. Black lines represent the maximum and minimum errors. Red dashed line represents the intrinsic stochastic error. The purple dot represents the average error value, the error bar represents the standard deviation of the error. The blue dashed line is the piecewise linear approximation. }
\label{X6}
\end{figure} 

\section{Conclusion}
In our analysis we observed the effect of different measurement error on the final predictions of the Deffuant model. The model shows an intrinsic error of 5.4 \% which does not depend on the number of agents or running time. Instead, this value is weakly dependent on the two model’s parameters $\varepsilon$ and $\mu$.

Regarding measurement errors, we observed that the model is independent on the number of levels of the scale if this number is above 3. However, if a scale with 3 or 2 levels is used the prediction error will quickly rise up to an average error of 34 \%. As most scales have more than three bins (especially multi-items scales), binning is likely to have little effect on predictions. This also shows a type of data incompatibility between the Deffuant model and binary ones such as the voter model. Indeed, these models are designed to work with 2-points scales, where the Deffuant model will have extremely high prediction error.

Random noise and distortions, contrary to binning, may have a massive effect on the model’s prediction. Specifically, the prediction error is roughly linear with both effects. However, random noise reaches a maximum of 40 \% for $\sigma = 0.5$ and then stops increasing. Instead, distortions keep contributing to the prediction error up to an average error bigger than 80 \%. Furthermore, by using values from scales obtained from the Wellcome Global Monitor, we found that here random noise produces a prediction error of 8.5 \% while distortion produces an error of 18 \%. This means that researchers interested in using real-world data should be particularly careful about these two effects.

One of the main difficulties when dealing with psychometrics scale is that it is not possible to directly measure noise and distortions \citep{Tversky1}. Indeed, we can directly identify the number of levels in one scale, by observing the number of possible scores. Instead, to observe the effect of noise and distortions we should compare the scale to a reference one, as done in figure 1. 

In practical terms this means that, if we only use one scale, the worst prediction error we would expect is the intrinsic stochastic error. However, if another person uses a different scale, her predictions may be totally different from the ones we obtained. Indeed, since the two scales are different, we can have any level of noise and distortions. 

This also means that if one study shows that the Deffuant model makes good predictions on a specific scale of political opinion, this does not mean it would work on every scale of political opinion. Indeed, a different scale may have strong distortions respect to the first one and so produce strongly different predictions.

Currently, the only possible way to compare the result of two different scales consist in directly comparing the scales themselves and checking the amount of noise and distortions. If these levels are small enough, the two predictions would be comparable with a reasonable uncertainty. However, further works may also focus on new ways of making the models more adaptable or more stable to measurement errors. For example, the Axelrod model of cultural diffusion do not rely on distances or arithmetic operations, which can make this type of models more robust to distortions \citep{Axelrod, Carpentras}

This result shows the importance of measurement in opinion dynamics literature. Indeed, until now, most of the analysis was dedicated to examining the model properties without considering the effect of using real-world data. However, here we show that for producing reliable predictions, we should also take into account the effect of measurement errors as they can massively affect predictions. Thus, we hope that in the future models will be tested also against these type of measurement distortions.

\section*{Acknowledgements}
This project has received funding from the European Union’s Horizon 2020 research and innovation programme under the Marie Skłodowska-Curie grant agreement No 891347 and from the European Research Council (ERC) under the European Union's Horizon 2020 research and innovation programme (grant agreement No. 802421).




% ...





 
\bibliographystyle{jasss}
\bibliography{bibliography.bib} 

\begin{thebibliography}{44}
\providecommand{\natexlab}[1]{#1}
\providecommand{\url}[1]{\texttt{#1}}
\providecommand{\urlprefix}{URL }
\expandafter\ifx\csname urlstyle\endcsname\relax
  \providecommand{\doi}[1]{doi:\discretionary{}{}{}#1}\else
  \providecommand{\doi}{doi:\discretionary{}{}{}\begingroup
  \urlstyle{rm}\Url}\fi

\bibitem[{Axelrod(1997)}]{Axelrod}
Axelrod, R. (1997).
\newblock The dissemination of culture: A model with local convergence and
  global polarization.
\newblock \textit{Journal of conflict resolution}, \textit{41}(2), 203--226

\bibitem[{Baggaley \& Hull(1983)}]{Baggaley}
Baggaley, A.~R. \& Hull, A.~L. (1983).
\newblock The effect of nonlinear transformations on a likert scale.
\newblock \textit{Evaluation \& the health professions}, \textit{6}(4),
  483--491

\bibitem[{Borsboom \& Scholten(2008)}]{Borsboom}
Borsboom, D. \& Scholten, A.~Z. (2008).
\newblock The rasch model and conjoint measurement theory from the perspective
  of psychometrics.
\newblock \textit{Theory \& Psychology}, \textit{18}(1), 111--117

\bibitem[{Carpentras et~al.(2020)Carpentras, Dinkelberg \& Quayle}]{Carpentras}
Carpentras, D., Dinkelberg, A. \& Quayle, M. (2020).
\newblock A new degree of freedom for opinion dynamics models: the
  arbitrariness of scales.
\newblock \textit{arXiv preprint arXiv:2010.04788}

\bibitem[{Castellano et~al.(2009)Castellano, Fortunato \&
  Loreto}]{Castellano_2009}
Castellano, C., Fortunato, S. \& Loreto, V. (2009).
\newblock Statistical physics of social dynamics.
\newblock \textit{Reviews of Modern Physics}, \textit{81}(2), 591–646.
\newblock \doi{10.1103/revmodphys.81.591}
\newline\urlprefix\url{http://dx.doi.org/10.1103/revmodphys.81.591}

\bibitem[{Coates et~al.(2018)Coates, Han \& Kleerekoper}]{Coates}
Coates, A., Han, L. \& Kleerekoper, A. (2018).
\newblock A unified framework for opinion dynamics.
\newblock In \textit{Proceedings of the 17th International Conference on
  Autonomous Agents and Multiagent Systems}. International Foundation for
  Autonomous Agents and Multiagent Systems

\bibitem[{Deffuant et~al.(2000)Deffuant, Neau, Amblard \&
  Weisbuch}]{Deffuant2000}
Deffuant, G., Neau, D., Amblard, F. \& Weisbuch, G. (2000).
\newblock Mixing beliefs among interacting agents.
\newblock \textit{Advances in Complex Systems}, \textit{3}(01n04), 87--98

\bibitem[{DeVellis(2016)}]{DeVellis}
DeVellis, R.~F. (2016).
\newblock \textit{Scale development: Theory and applications}, vol.~26.
\newblock Sage publications

\bibitem[{Dong et~al.(2018)Dong, Zhan, Kou, Ding \& Liang}]{Dong}
Dong, Y., Zhan, M., Kou, G., Ding, Z. \& Liang, H. (2018).
\newblock A survey on the fusion process in opinion dynamics.
\newblock \textit{Information Fusion}, \textit{43}, 57--65

\bibitem[{Duggins(2017)}]{Duggins}
Duggins, P. (2017).
\newblock A psychologically-motivated model of opinion change with applications
  to american politics.
\newblock \textit{Journal of Artificial Societies and Social Simulation},
  \textit{20}(1), 13.
\newblock \doi{10.18564/jasss.3316}
\newline\urlprefix\url{http://dx.doi.org/10.18564/jasss.3316}

\bibitem[{Everett(2013)}]{Everett}
Everett, J.~A. (2013).
\newblock The 12 item social and economic conservatism scale (secs).
\newblock \textit{PloS one}, \textit{8}(12), e82131

\bibitem[{Feir \& Toothaker(1974)}]{Feir}
Feir, B.~J. \& Toothaker, L.~E. (1974).
\newblock The anova f-test versus the kruskal-wallis test: A robustness study.

\bibitem[{Fennell et~al.(2021)Fennell, Burke, Quayle \& Gleeson}]{Fennell}
Fennell, S.~C., Burke, K., Quayle, M. \& Gleeson, J.~P. (2021).
\newblock Generalized mean-field approximation for the deffuant opinion
  dynamics model on networks.
\newblock \textit{Physical Review E}, \textit{103}(1), 012314

\bibitem[{Flache et~al.(2017)Flache, M\"{a}s, Feliciani, Chattoe-Brown,
  Deffuant, Huet \& Lorenz}]{Flache}
Flache, A., M\"{a}s, M., Feliciani, T., Chattoe-Brown, E., Deffuant, G., Huet,
  S. \& Lorenz, J. (2017).
\newblock Models of social influence: Towards the next frontiers.
\newblock \textit{Journal of Artificial Societies and Social Simulation},
  \textit{20}(4), 2.
\newblock \doi{10.18564/jasss.3521}
\newline\urlprefix\url{http://dx.doi.org/10.18564/jasss.3521}

\bibitem[{Glass et~al.(1972)Glass, Peckham \& Sanders}]{Glass}
Glass, G.~V., Peckham, P.~D. \& Sanders, J.~R. (1972).
\newblock Consequences of failure to meet assumptions underlying the fixed
  effects analyses of variance and covariance.
\newblock \textit{Review of educational research}, \textit{42}(3), 237--288

\bibitem[{Halliday et~al.(2013)Halliday, Resnick \& Walker}]{Halliday}
Halliday, D., Resnick, R. \& Walker, J. (2013).
\newblock \textit{Fundamentals of physics}.
\newblock John Wiley \& Sons

\bibitem[{Heene(2013)}]{Heene}
Heene, M. (2013).
\newblock Additive conjoint measurement and the resistance toward
  falsifiability in psychology.
\newblock \textit{Frontiers in psychology}, \textit{4}, 246

\bibitem[{Heisenberg(1985)}]{Heisenberg}
Heisenberg, W. (1985).
\newblock {\"U}ber den anschaulichen inhalt der quantentheoretischen kinematik
  und mechanik

\bibitem[{Holley \& Liggett(1975)}]{Holley}
Holley, R.~A. \& Liggett, T.~M. (1975).
\newblock Ergodic theorems for weakly interacting infinite systems and the
  voter model.
\newblock \textit{The annals of probability}, (pp. 643--663)

\bibitem[{Howarth(2006)}]{Howarth}
Howarth, C. (2006).
\newblock How social representations of attitudes have informed attitude
  theories: the consensual and the reified.
\newblock \textit{Theory \& Psychology}, \textit{16}(5), 691--714

\bibitem[{Innes(2014)}]{Innes}
Innes, C.~R. (2014).
\newblock \textit{Quantifying the effect of open-mindedness on opinion dynamics
  and advertising optimization}.
\newblock Ph.D. thesis, Science:

\bibitem[{Jia et~al.(2015)Jia, MirTabatabaei, Friedkin \& Bullo}]{Jia}
Jia, P., MirTabatabaei, A., Friedkin, N.~E. \& Bullo, F. (2015).
\newblock Opinion dynamics and the evolution of social power in influence
  networks.
\newblock \textit{SIAM Review}, \textit{57}(3), 367--397.
\newblock \doi{10.1137/130913250}
\newline\urlprefix\url{http://dx.doi.org/10.1137/130913250}

\bibitem[{Krantz et~al.(2006{\natexlab{a}})Krantz, Luce, Suppes \&
  Tversky}]{Tversky1}
Krantz, D., Luce, D., Suppes, P. \& Tversky, A. (2006{\natexlab{a}}).
\newblock \textit{Foundations of measurement, Vol. 1: Additive and polynomial
  representations}.
\newblock Dover Publications

\bibitem[{Krantz et~al.(2006{\natexlab{b}})Krantz, Luce, Suppes \&
  Tversky}]{Tversky2}
Krantz, D., Luce, D., Suppes, P. \& Tversky, A. (2006{\natexlab{b}}).
\newblock \textit{Foundations of measurement, Vol. 2: Geometrical, Threshold,
  and Probabilistic Representations}.
\newblock Dover Publications

\bibitem[{Krantz et~al.(2006{\natexlab{c}})Krantz, Luce, Suppes \&
  Tversky}]{Tversky3}
Krantz, D., Luce, D., Suppes, P. \& Tversky, A. (2006{\natexlab{c}}).
\newblock \textit{Foundations of measurement, Vol. 3: Representation,
  Axiomatization, and Invariance}.
\newblock Dover Publications

\bibitem[{Landau \& Lifshitz(2013)}]{Landau}
Landau, L.~D. \& Lifshitz, E.~M. (2013).
\newblock \textit{Quantum mechanics: non-relativistic theory}, vol.~3.
\newblock Elsevier

\bibitem[{Lantz(2013)}]{Lanz}
Lantz, B. (2013).
\newblock Equidistance of likert-type scales and validation of inferential
  methods using experiments and simulations.
\newblock \textit{The Electronic Journal of Business Research Methods},
  \textit{11}(1), 16--28

\bibitem[{Lean et~al.(2021)Lean, H{\'o}lm, Bonavita, Bormann, McNally \&
  J{\"a}rvinen}]{Lean}
Lean, P., H{\'o}lm, E., Bonavita, M., Bormann, N., McNally, A. \& J{\"a}rvinen,
  H. (2021).
\newblock Continuous data assimilation for global numerical weather prediction.
\newblock \textit{Quarterly Journal of the Royal Meteorological Society},
  \textit{147}(734), 273--288

\bibitem[{Leuthold(1975)}]{Leuthold}
Leuthold, R.~M. (1975).
\newblock On the use of theil's inequality coefficients.
\newblock \textit{American Journal of Agricultural Economics}, \textit{57}(2),
  344--346

\bibitem[{Luce \& Tukey(1964)}]{Luce64}
Luce, R.~D. \& Tukey, J.~W. (1964).
\newblock Simultaneous conjoint measurement: A new type of fundamental
  measurement.
\newblock \textit{Journal of mathematical psychology}, \textit{1}(1), 1--27

\bibitem[{MSCA-IF(2020)}]{Dynamod}
MSCA-IF (2020).
\newblock Dynamod-vaccine-data.
\newblock \url{https://cordis.europa.eu/project/id/891347}

\bibitem[{Nering \& Ostini(2011)}]{IRT}
Nering, M.~L. \& Ostini, R. (2011).
\newblock \textit{Handbook of polytomous item response theory models}.
\newblock Taylor \& Francis

\bibitem[{Nunnally \& Bernstein(1994)}]{Nunnally}
Nunnally, J.~C. \& Bernstein, I.~H. (1994).
\newblock Psychometric theory

\bibitem[{Sayood(2005)}]{Sayood}
Sayood, K. (2005).
\newblock \textit{Introduction to Data Compression, 3rd}.
\newblock Morgan Kaufmann

\bibitem[{Smith et~al.(1997)}]{Smith}
Smith, S.~W. et~al. (1997).
\newblock \textit{The scientist and engineer's guide to digital signal
  processing}, vol.~14.
\newblock California Technical Pub. San Diego

\bibitem[{Soskind et~al.(2014)Soskind, Walvick, Giranda, Laslo \&
  Gifford}]{Soskind}
Soskind, Y., Walvick, R., Giranda, C., Laslo, D. \& Gifford, R. (2014).
\newblock Point-spread function-based characterization of optical systems.
\newblock In \textit{Photonic Instrumentation Engineering}, vol. 8992, (p.
  89920E). International Society for Optics and Photonics

\bibitem[{Stevens(1946)}]{Stevens_1946}
Stevens, S.~S. (1946).
\newblock On the theory of scales of measurement.
\newblock \textit{Science}, \textit{103}(2684), 677–680.
\newblock \doi{10.1126/science.103.2684.677}
\newline\urlprefix\url{http://dx.doi.org/10.1126/science.103.2684.677}

\bibitem[{Stevens(1951)}]{Stevens51}
Stevens, S.~S. (1951).
\newblock \textit{Mathematics, measurement, and psychophysics.}
\newblock Wiley

\bibitem[{Taylor(1997)}]{Taylor}
Taylor, J. (1997).
\newblock \textit{Introduction to error analysis, the study of uncertainties in
  physical measurements}.
\newblock Univ Science Books

\bibitem[{Ten~Broeke et~al.(2016)Ten~Broeke, Van~Voorn \& Ligtenberg}]{Broeke}
Ten~Broeke, G., Van~Voorn, G. \& Ligtenberg, A. (2016).
\newblock Which sensitivity analysis method should i use for my agent-based
  model?
\newblock \textit{Journal of Artificial Societies and Social Simulation},
  \textit{19}(1)

\bibitem[{ToRealSim(2018)}]{ToRealSim}
ToRealSim (2018).
\newblock Towards realistic computational models of social influence dynamics.
\newblock \url{https://gtr.ukri.org/projects?ref=ES\%2FS015159\%2F1}

\bibitem[{Valori et~al.(2012)Valori, Picciolo, Allansdottir \&
  Garlaschelli}]{Valori}
Valori, L., Picciolo, F., Allansdottir, A. \& Garlaschelli, D. (2012).
\newblock Reconciling long-term cultural diversity and short-term collective
  social behavior.
\newblock \textit{Proceedings of the National Academy of Sciences},
  \textit{109}(4), 1068--1073

\bibitem[{Vazquez \& Egu{\'\i}luz(2008)}]{Vazquez}
Vazquez, F. \& Egu{\'\i}luz, V.~M. (2008).
\newblock Analytical solution of the voter model on disordered networks.
\newblock \textit{arXiv preprint arXiv:0803.1686}

\bibitem[{WellcomeTrust(2018)}]{WGM}
WellcomeTrust (2018).
\newblock Wellcome global monitor.
\newblock \url{https://wellcome.org/reports/wellcome-global-monitor/2018}

\end{thebibliography}


\end{document}